\documentclass[a4paper,11pt, french]{article}

\usepackage{color}
\usepackage{babelfr,helvet}
\usepackage{graphicx}
\usepackage{graphics}
\usepackage{amssymb}
\usepackage{pdfpages}
\usepackage[authoryear]{natbib}
\usepackage{url}
\usepackage{titlesec}
\usepackage[small,hang]{caption}
\begin{document}
\makeatletter
\renewcommand\section{\@startsection {section}{1}{\z@}%
                                   {-3.5ex \@plus -1ex \@minus -.2ex}%
                                   {2.3ex \@plus.2ex}%
                                   {\normalfont\normalsize\bfseries}}% from \Large
\renewcommand\subsection{\@startsection{subsection}{2}{\z@}%
                                     {-3.25ex\@plus -1ex \@minus -.2ex}%
                                     {1.5ex \@plus .2ex}%
                                     {\normalfont\normalsize\bfseries}}% from \large

\renewcommand{\captionfont}{\normalfont \small}
\makeatother
\begin{center}
%\begin{LARGE}
{{\textbf{\large{The Gaia-FUN-SSO observation campaign of 99942 Apophis: A preliminary test for the network}}}} \\
\medskip
\small{D. Bancelin$^{1,2}$,  W. Thuillot$^2$, A. Ivantsov$^3$, J. Desmars$^{4,2}$, M. Assafin$^5$, S. Eggl$^2$, D.
Hestroffer$^2$, P. Rocher$^2$, B. Carry$^2$, P. David$^2$ and the Gaia-FUN-SSO team} \\ 
\medskip
1. Institute of Astrophysics, University of Vienna, Austria; david.bancelin@univie.ac.at \\ 2. IMCCE, Paris
observatory, 75014 Paris, France \\ 3.
Faculty of Aerospace Engineering, Haifa, Isra\"el \\ 4. Observatorio National, Rio de Janeiro, Brazil \\ 5. Observatorio
do Valongo, Rio de Janeiro, Brazil \\
%\end{LARGE}
\end{center}

\bigskip

\section{Abstract}

 In order
to test the coordination and evaluate the overall performance of the Gaia-FUN-SSO, an observation campaign on the
Potentially Hazardous Asteroid (99 942) Apophis was conducted from 12/21/2012 to 5/2/2013 providing 2732 high
quality astrometric observations. We show that a consistent reduction of astrometric campaigns with reliable
stellar catalogs substantially improves the quality of astrometric results. We present evidence that the new data will
help to reduce the orbit uncertainty of Apophis during its close approach in 2029.

\section{Introduction}

In the framework of the Gaia mission, an alert mode (a ground-based follow-up network \citep{thuillot11}), has been set
up in order to identify newly detected objects and trigger complementary observations from the ground, since the
satellite cannot keep monitoring its discoveries. Specific training campaigns have been organized during the past three
years. In particular, the observation campaign of Apophis from 12/21/2012 to 5/2/2013, providing 2732 valuable
astrometric measurements among the collection of extensive observations. Some of the observations performed,
already submitted to the MPC, have been
reduced by
the observers themselves, using their preferred tools and astrometric catalogs. However, we decided to conduct a
complementary homogeneous reduction, with all CCD images recorded during this campaign
using the PRAIA reduction pipeline \citep{assafin11} and the UCAC4 astrometric catalog \citep{zacharias13}. This yields
to consistent set of 2732 astrometric measurements of Apophis. In the following we will discuss data analysis of the
observations acquired by the Gaia-FUN-SSO. We will show that a consistent analysis can decrease
systematic errors and boost the quality of astrometric positions.

\section{ Data analysis}

Among the 2732 astrometric measurements, 629 had already been sent ot the MPC by the observers. This gives us an unique
opportunity to compare the consistency of these observations according to the catalog used for the data reduction. We
thus
define:

\begin{itemize}
 \item D$_{\scriptscriptstyle {MPC}}$ as the 629 duplicated Gaia-FUN-SSO astrometric measurements already sent to the
MPC by the observers.
The corresponding observations were reduced with various astrometric software packages and catalogs.
\item D$_{\scriptscriptstyle {PRAIA}}$ as the same 629 Gaia-FUN-SSO observations, but re-reduced with PRAIA using the
UCAC4 astrometric
catalog.
\item S$_{\scriptscriptstyle {NEW}}$ as the 2109 unsent observations.

\end{itemize}

\subsection{Alert and recovery process}
%  \begin{center}
%    \textbf{Alert and recovery process}
%   \end{center}
  
\begin{figure}[h!]
\centering
\includegraphics[width= 8cm]{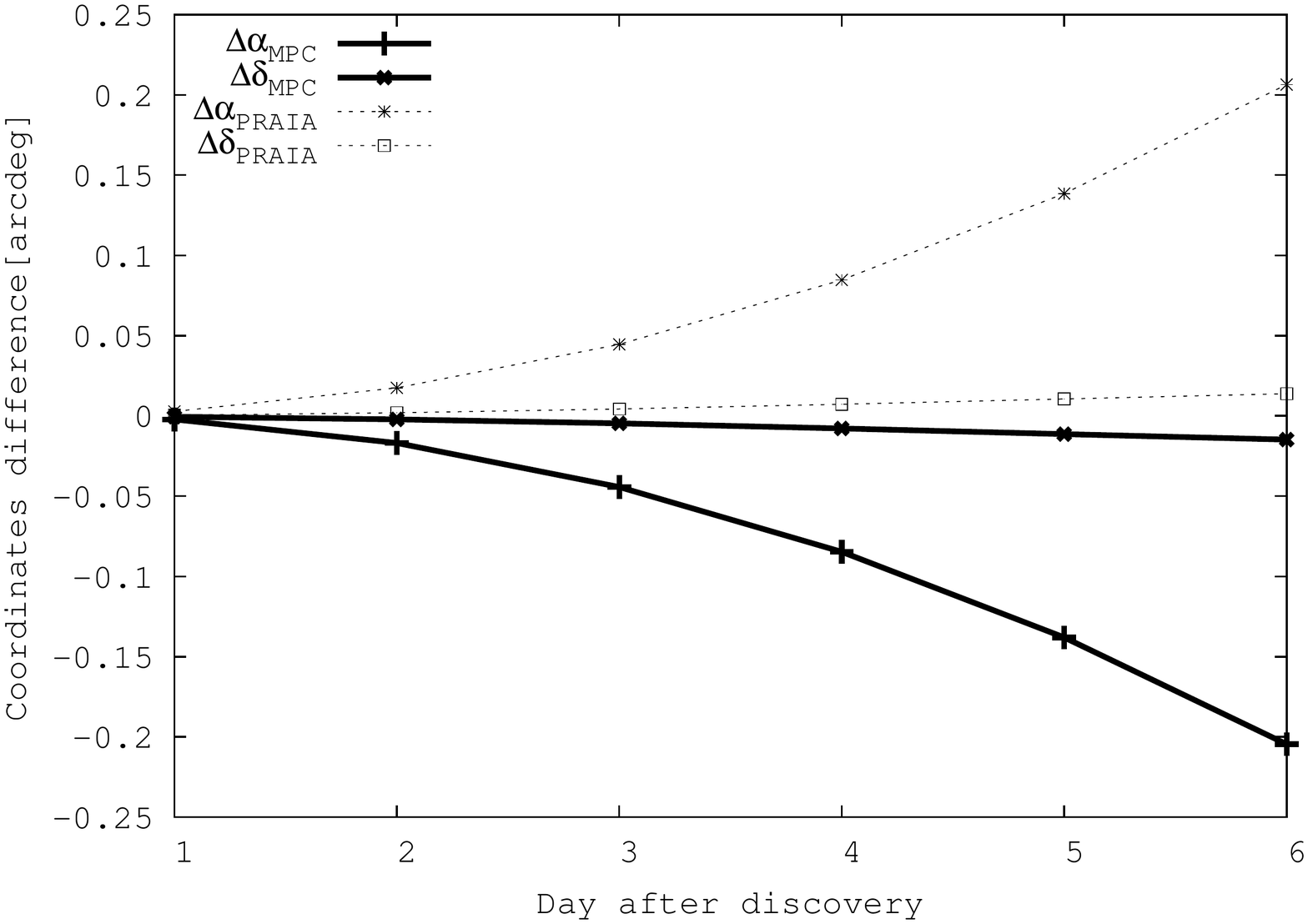}\\
\caption{\small{The graph shows the time evolution of the coordinate differences
($\Delta\alpha$,$\Delta\delta$)$_{\scriptscriptstyle {MPC}}$ and ($\Delta\alpha$,$\Delta\delta$)$_{\scriptscriptstyle
{PRAIA}}$ between orbit solutions derived from different data sets with respect to the nominal solution (obtained using
all the optical and radar data available)}}
\label{F:evolution}
\end{figure}

Using a similar approach as \cite{bancelin12b}, we aim to assess how far
the predicted position
can drift from the real one in a given amount of time. Let us consider a hypothetical discovery of an asteroid during
the Gaia-FUN-SSO campaign. We will use the observational data of Apophis, but we shall assume its orbit was previously
unknown. Furthermore, we assume that the hypothetical discovery has happened on the first night recorded in the
duplicated measurements D$_{\scriptscriptstyle {PRAIA}}$ and D$_{\scriptscriptstyle {MPC}}$. This first night set is
used to determine the orbit and orbital
elements covariance matrix of the new object. We then propagated the orbit solutions and uncertainties obtained from
both
sets up to six days after the discovery. One week after the discovery the coordinate differences $\Delta\alpha$ and
$\Delta\delta$ between D$_{\scriptscriptstyle {PRAIA}}$ , D$_{\scriptscriptstyle {MPC}}$ and the "true" position of
Apophis (obtained with the 2004-2014 optical and radar data) are evaluated. Figure \ref{F:evolution} shows how the
differences in astrometric coordinates evolve for both sets of
measurements during the six days following the discovery. The opposing orientation of the
($\Delta\alpha$,$\Delta\delta$)$_{\scriptscriptstyle {MPC}}$ and
($\Delta\alpha$,$\Delta\delta$)$_{\scriptscriptstyle {PRAIA}}$ curves is due to the different preliminary orbital
elements
found using D$_{PRAIA}$ and D$_{\scriptscriptstyle {MPC}}$ . One can see that
($\Delta\alpha$,$\Delta\delta$)$_{\scriptscriptstyle {MPC}}$ and
($\Delta\alpha$,$\Delta\delta$)$_{\scriptscriptstyle {PRAIA}}$ are of the same order of magnitude. Consequently, the
method of data reduction is unlikely to have a significant impact on the recovery process within the network.

\subsection{Position uncertainty propagation for new discoveries}
%   \begin{center}
%    \textbf{Position uncertainty propagation for new discoveries}
%   \end{center}

\begin{figure}[h!]
\centering
\includegraphics[width= 8cm]{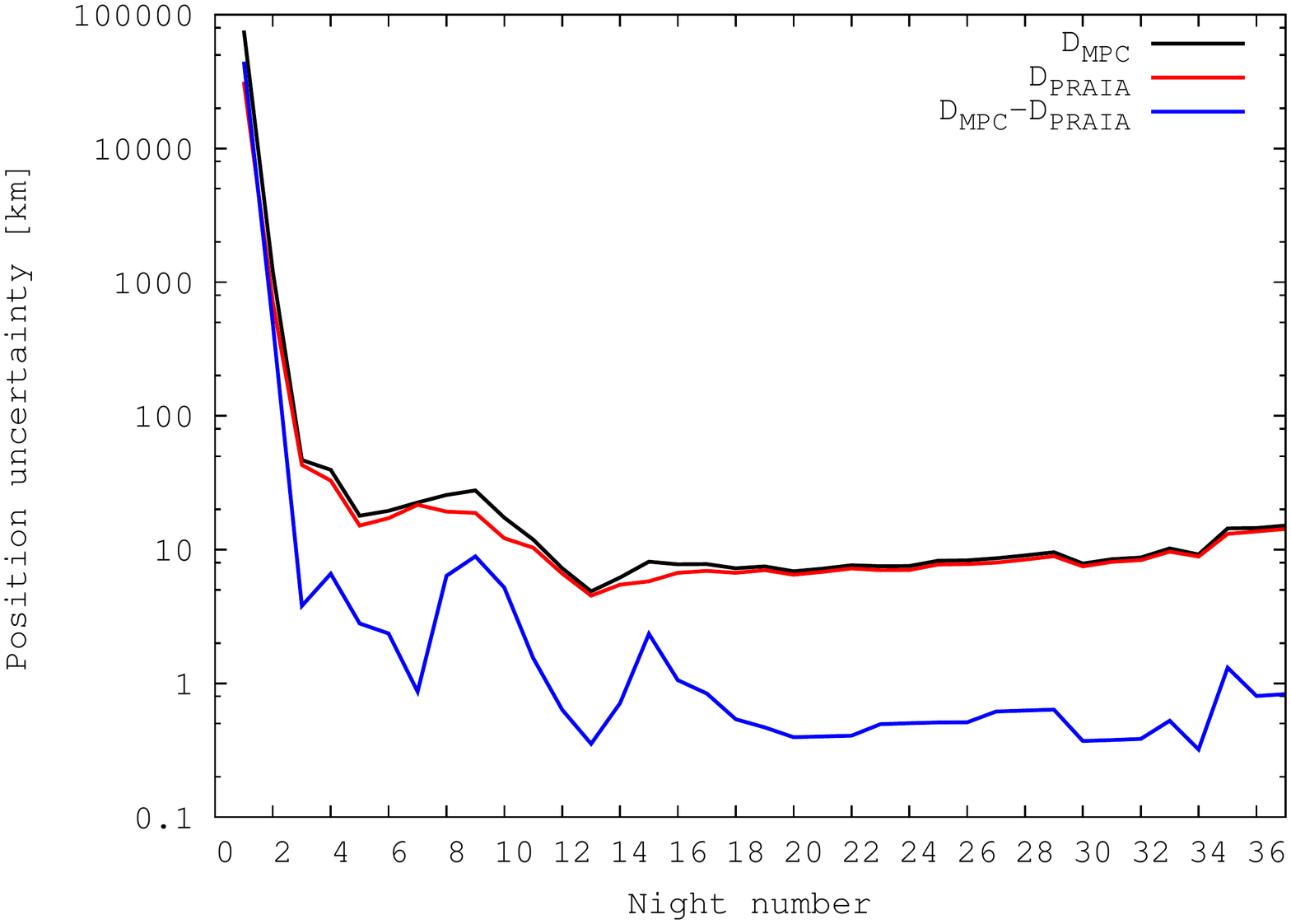}\\
\caption{\small{Geocentric position uncertainty evolution as a function of the number of observation nights for the
duplicated measurement. The difference between the sets D$_{\scriptscriptstyle {MPC}}$ - D$_{\scriptscriptstyle
{PRAIA}}$ is also indicated}}
\label{F:uncertainty}
\end{figure}

We are now interested in how the position uncertainty evolves when more observations become available during the nights
following an asteroid's discovery. As we assume the asteroid to be newly discovered, a preliminary orbit determination
is conducted after the first night of the sets D$_{\scriptscriptstyle {PRAIA}}$ and D$_{\scriptscriptstyle {MPC}}$ and
an orbital improvement is performed. Uncertainties on the geocentric position is then calculated. This allows us to
compare the impact of the reduction pipeline on the uncertainty evolution of a newly found object. Figure
\ref{F:uncertainty} shows that at the discovery night (first night), uncertainties are large for both sets. However, it
is only after the 10$^{\scriptscriptstyle {th}}$ night that the difference D$_{\scriptscriptstyle {MPC}}$ -
D$_{\scriptscriptstyle {PRAIA}}$ drops permanently below 10 km. Since between the first and the 10$^{\scriptscriptstyle
{th}}$ night span an arc of 26 days, there is a real advantage in consistent reduction regarding the position
uncertainty propagation of follow up campaigns.

\subsection{Orbit propagation for Apophis}
%  \begin{center}
%    \textbf{Orbit propagation for Apophis}
%   \end{center}
  
We will now proceed to study whether orbits and initial uncertainties constructed from different sets of observations
can cause a significant change in the propagated uncertainties of Apophis' orbit. The process then works as follows.
After an initial orbit determination, an orbit adjustment based on a differential correction is performed. This results
in the uncertainties of the asteroids orbit in form of an orbital element covariance matrix. The resulting uncertainties
can then be propagated to the 2029 b-plane and its long axis was used to indicate the 1$\sigma$ uncertainty value. A
quick first check can be performed using the duplicated measurement sets D$_{\scriptscriptstyle {MPC}}$ and
D$_{\scriptscriptstyle {PRAIA}}$. The propagated uncertainty with D$_{\scriptscriptstyle {PRAIA}}$ improves the
1$\sigma$ uncertainty obtaines with D$_{\scriptscriptstyle {MPC}}$ by $\sim$ 14\% which is non
negligible for the impact probability assessment with short arc data.

\subsection{ Impact of Gaia-FUN-SSO Observations on Orbit Uncertainties}

Our aim is to investigate whether the consistent data produced during the Gaia-FUN-SSO campaign can impact orbital
solutions and b-plane uncertainties through the example of Apophis. To this end we compare orbits and uncertainties
derived from five observational data sets:

\begin{itemize}
 \item S$_{\scriptscriptstyle {1}}$ = [2004-2014]$_{\scriptscriptstyle {MPC}}$ + radar
 \item S$_{\scriptscriptstyle {2}}$ = [2004-2014]$_{\scriptscriptstyle {MPC}}$ - D$_{\scriptscriptstyle {MPC}}$ +
D$_{\scriptscriptstyle {PRAIA}}$ + radar 
 \item S$_{\scriptscriptstyle {3}}$ = S$_{\scriptscriptstyle {1}}$ + S$_{\scriptscriptstyle {\scriptscriptstyle {NEW}}}$
 \item S$_{\scriptscriptstyle {4}}$ = S$_{\scriptscriptstyle {2}}$ + S$_{\scriptscriptstyle {NEW}}$
 \item S$_{\scriptscriptstyle {5}}$ =  S$_{\scriptscriptstyle {NEW}}$ + D$_{\scriptscriptstyle {PRAIA}}$ + radar

\end{itemize}

where [2004-2014]$_{\scriptscriptstyle {MPC}}$ refers to the 4138 optical data as present in the MPC database. We
propagated each nominal orbit resulting from the individual sets of observations together with its covariances up to
2029 where we evaluated the position uncertainties projected onto the b-plane. Table \ref{T:khi} summarizes the quality
of the orbital fit and the
2029 b-plane uncertainty resulting from the orbit propagation. The presented results suggests the sets containing
D$_{\scriptscriptstyle {PRAIA}}$ instead of D$_{\scriptscriptstyle {MPC}}$ result in smaller uncertainties in Apophis'
positions in the 2029 b-plane. Indeed, even for a well-known orbit (with a 10-years arc data length), both optical and
radar $\chi^2$ values show better results when D$_{\scriptscriptstyle {PRAIA}}$ measurements are used. Hence, we
speculate that current orbit solutions of NEAs can be improved using consistent data. Furthermore, consistent data
reduction with a good astrometric catalog can also result in smaller uncertainties in the b-plane coordinates of PHAs,
as was shown for the 2029-b-plane of Apophis. Moreover, we see that the Gaia-FUN-SSO observations and radar data
(S$_{\scriptscriptstyle
{5}}$) suffice to produce b-plane uncertainty values that are very close to those sets that contain all available
observations.

%-------------------------------------------------------------------------------
\begin{table}
 \begin{center}
  \caption{Orbital accuracy information -- fit residuals and b-plane uncertainty -- computed with different sets of
observations. We also computed the difference in b-plane distance $\Delta_{\scriptscriptstyle {i}}$ for each set with
respect to the distance $\Delta_{\scriptscriptstyle {1}}$ obtained from S$_{\scriptscriptstyle {1}}$}
  \label{T:khi}
  \begin{tabular}{|c|c|c|c|c|}
   \hline
   \hline
    & $\chi_{\scriptscriptstyle {opt}}^2$  &  $\chi_{\scriptscriptstyle {rad}}^2$ & $\sigma_{\xi}$ &
$\Delta_{\scriptscriptstyle {i}}$ - $\Delta_{\scriptscriptstyle {1}}$ \\
    &                 &                 & [km] & [km] \\
  \hline
S$_{\scriptscriptstyle {1}}$ & 0.227& 0.434 & 2.99  & 0 \cr
\hline
S$_{\scriptscriptstyle {2}}$ & 0.224&0.426  &2.94 & 0  \cr
\hline
S$_{\scriptscriptstyle {3}}$ & 0.157& 0.175&2.45  &1.5 \cr
\hline
S$_{\scriptscriptstyle {4}}$ & 0.155& 0.174&2.43 & 1.5\cr
\hline
S$_{\scriptscriptstyle {5}}$ & 0.021 & 0.095&3.24 & 3 \cr

\hline
\hline
  \end{tabular}
 \end{center}
\end{table}
%-------------------------------------------------------------------------------

\section{Conclusion}

 A large amount of astrometric data was collected  during the latest period of observability of Apophis in
2012-2013 and processed in a homogeneous fashion using the PRAIA
reduction software and the UCAC4 catalog data. Using the 629 duplicated data from the 2732 precise
astrometric measurements provided by 19 observatories, we could show that the
recovery process of new objects when their observational data arcs span less than one night won't be impact when
considering MPC or PRAIA data. However, a consistent data reduction of a newly discovered asteroids during this
observation campaign would have led to a greater reduction of NEO position uncertainties. Finally, the example of
Apophis reveals that, even for well-known orbits, the use of consistent data can improve the current $\chi^2$ of both
optical and radar data.

\bibliographystyle{plainnat}
\bibliography{biblio}

\end{document}